\documentclass[a4paper,11pt]{article}
\usepackage{pos}
\usepackage{subfigure}
\usepackage{customCommands}

\title{Optimizing the Wasserstein GAN for TeV Gamma Ray Detection with VERITAS}
 \ShortTitle{GAN for TeV Gamma Ray Detection}

\author*[a,b]{Deivid Ribeiro}\onbehalf{ for the VERITAS Collaboration}
\author[a]{Yuping Zheng}
\author[a]{Ramana Sankar}
\author[a]{Kameswara Mantha}

\affiliation[a]{School of Physics \& Astronomy, 116 Church St SE, Minneapolis, USA}
\affiliation[b]{Minnesota Institute for Astrophysics, 116 Church St SE, Minneapolis, USA}

\emailAdd{ribei056@umn.edu}

\abstract{The observation of very-high-energy (VHE, E>100 GeV) gamma rays is mediated by the imaging atmospheric Cherenkov technique (IACTs). At these energies, gamma rays interact with the atmosphere to create a cascade of electromagnetic air showers that are visible to the IACT cameras on the ground with distinct morphological and temporal features. However, hadrons with significantly higher incidence rates are also imaged with similar features, and must be distinguished with handpicked parameters extracted from the images. The advent of sophisticated deep learning models has enabled an alternative image analysis technique that has been shown to improve the detection of gamma rays, by improving background rejection. In this study, we propose an unsupervised Wasserstein Generative Adversarial Network (WGAN) framework trained on normalized, uncleaned stereoscopic shower images of real events from the VERITAS observatory to extract the landscape of their latent space and optimize against the corresponding inferred latent space of simulated gamma-ray events. We aim to develop a data driven approach to guide the understanding of the extracted features of real gamma-ray images, and will optimize the WGAN to calculate a probabilistic prediction of “gamma-ness" per event. In this poster, we present results of ongoing work toward the optimization of the WGAN, including the exploration of conditional parameters and multi-task learning.
}

\ConferenceLogo{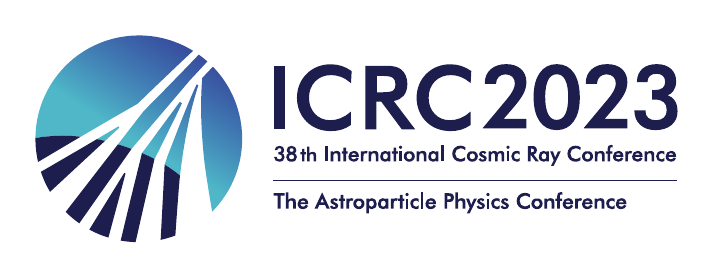}

\FullConference{%
38th International Cosmic Ray Conference (ICRC2023)\\
  26 July - 3 August, 2023\\
  Nagoya, Japan}

\begin{document}
\maketitle

\section{Introduction}
Gamma rays are the most energetic forms of  electromagnetic radiation and originate from the most extreme environments in the universe. We can detect gamma rays indirectly through the imaging atmospheric Cherenkov technique (IACTs)\cite{johnson2018thermal}. Gamma rays entering the atmosphere interact with particles and produce a cascade of secondary particles known as an Extensive Air Shower (EAS). These particles travel at speeds faster than the speed of light in air, producing a faint blue light called Cherenkov Radiation, which enables ground-based gamma-ray detection. However, the hadronic emission also produces an EAS. The ratio of a gamma-initiated shower to a hadron-initiated shower can range between 1:1000 and 1:10000 depending on the energy \cite{aharonian2008}. To detect the VHE gamma-ray signal, we want to find effective ways to distinguish the gamma ray and hadron EAS. This challenge has led to extensive research on gamma/hadron separation. 

In this work, we use observed data from the Very Energetic Radiation Imaging Telescope Array System (VERITAS). VERITAS is an array of four telescopes located at the Fred Lawrence Whipple Observatory in southern Arizona. VERITAS observes both gamma rays and hadronic background, where the gamma-ray initiated EAS and the hadron-initiated EAS are theoretically expected to produce different particles, leading to different shower evolutions. Images of the Cherenkov light emitted by both gamma ray and hadronic showers detected by ground-based observatories are expected to have distinct morphological and temporal features, however for  hadronic EAS dominated by electromagnetic interactions, there can be significant overlap in these features. The traditional gamma/hadron separation method uses Hillas parameters \cite{hillas1998spectrum}, which are parametric quantities that describe the image-level signal moments. However, Hillas parameter-based techniques do not take advantage of the full camera image of the EAS, since performance is optimized only on cleaned images devoid of background pixels. 

There is some existing work using supervised machine learning techniques where a labeled dataset generated through Monte-Carlo simulations is used to train the machine learning model (see \cite{capistran2015new,krause2017improved,spencer2021advanced}). However, the discrepancy between the real and simulated data potentially affects the resulting analysis, i.e., the model trained on the simulated data may not be general enough to capture the features of the real data such as features dependent on observed sky location and conditions.  Therefore, we propose to apply an unsupervised learning technique trained on the unlabeled, real observation dataset with the goal to discover a data-driven distinction between gamma and hadronic signals. The model is trained on the real dataset, and then simulated gamma-ray events and background datasets are used for the inference stage. 

The relatively low incidence of gamma rays with respect to the hadronic signals in ground based observed data renders the gamma/hadron separation problem to be an anomaly detection task, where anomalies are defined as the rare/poorly generalized samples within the data. We use a Generative Adversarial Neural Network (GAN) for anomaly detection which follows the ideas of the work on GAN-based anomaly detection in medical diagnosis (\cite{schlegl2019f}).


GANs were first proposed in \cite{Goodfellow2014GenerativeAN}, where two competing models work towards training each other. In a GAN, the two models are trained simultaneously: a generative model (generator) that captures the data distribution and a discriminative model (discriminator) that estimates the probability that a sample is true (from data distribution) or fake (from model distribution). The objective of the generator is to generate the best possible images, while the role of the discriminator is to force the generator to be generalized. The goal of a GAN is to find a balance between the generator and discriminator.

The original GAN model (or ``vanilla" GAN) suffers from training instability and mode collapse. Therefore, many variants of GANs have been developed to compensate for the shortcomings of the vanilla GAN. Wasserstein GANs (WGANs) use an improved metric, which is implemented by constraint clipping, in order to prevent mode collapse (which results in the generator producing grey noise images) \cite{Arjovsky2017WassersteinG}. However, this naive weight clipping results in unstable training. Therefore, the WGAN was augmented with a gradient penalty to stabilize the model (WGAN-gp,\cite{gulrajani2017improved}). In this project, we utilize WGAN-gp (hereafter just WGAN) to learn the feature representation of the real dataset due to its promising performance.

\section{Dataset Description} \label{sec:Dataset desciption}
The data from the VERITAS observations of the Crab Nebula source is used for training. The Crab Nebula was the first astrophysical gamma-ray source reliably detected from the ground and is one of the most extensively studied gamma-ray objects. It has produced some of the highest-energy photons ever detected, and is a stable source that is used for instrument calibration \cite{spencer2021advanced}.

Each of the four VERITAS imaging cameras uses 499 photomultiplier tubes (PMTs) arranged in a hexagonal pattern. Cubic interpolation is used to transform the hexagonal pixels into a $96\times96$ image with square pixels, which fits the deep learning framework better. There are no ground-truth labels for these images; the gamma-ray signal events are mixed with randomly triggered images (noise) and hadronic EAS images. Our dataset consists of $64,000$ events each with a dimension of $96\times96$ pixels across 4 dimensions, for each telescope. Simultaneous images from all four telescopes provides a stereoscopic view of the EAS, improving direction reconstruction and gamma/hadron separation. This setup allows us to train the deep learning model stereoscopically.

\begin{figure}[h]
\centering
\includegraphics[width=0.5\textwidth]{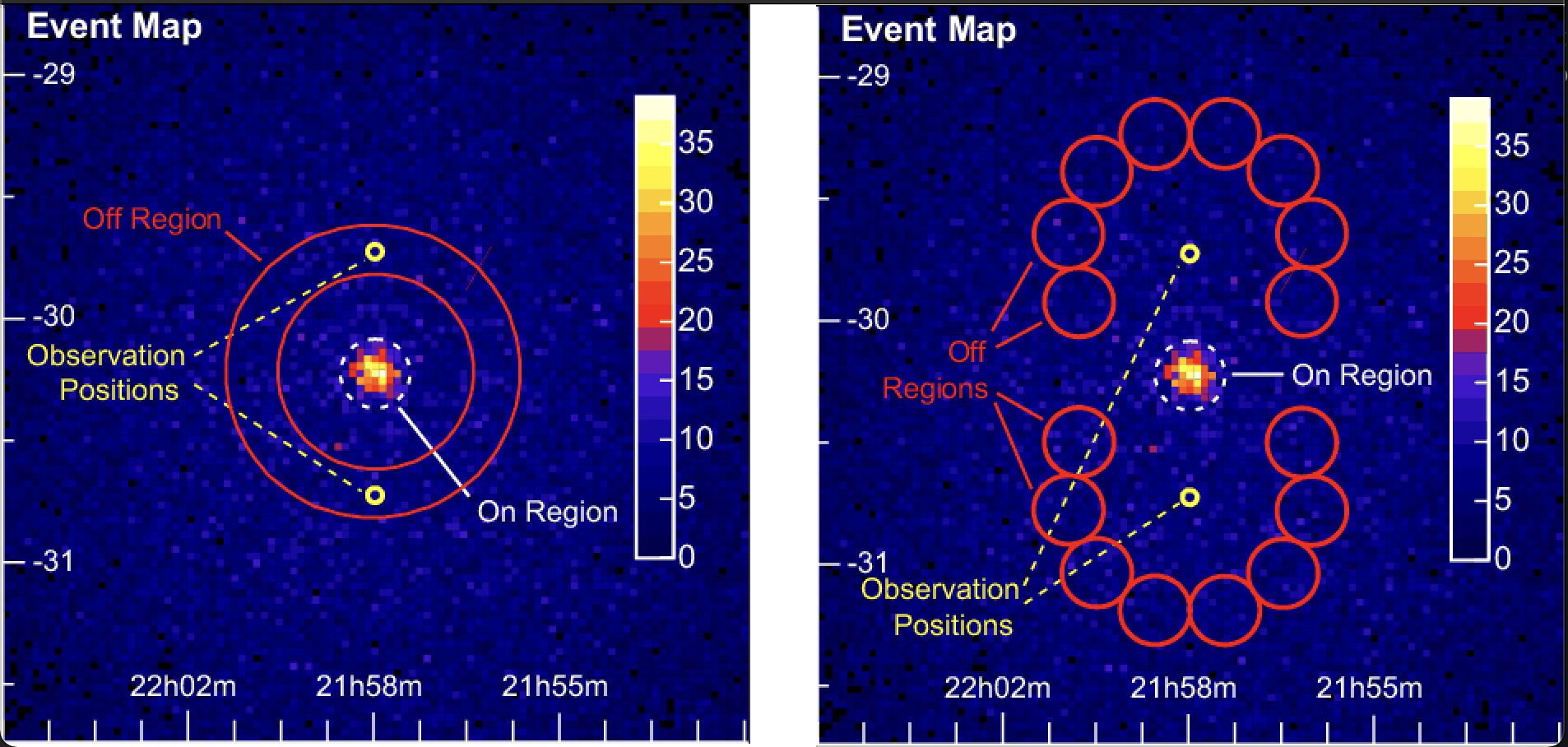}
\caption{Illustration of two background-region strategies. Left) Ring Background method (RB), where events in the off-region ring radially spread around the source are used for background. Right) Reflected region (RR), where background regions of the same size of the on-region are evenly spread around the observation positions \cite{berge2007a}.}
\label{fig:off-source}
\end{figure}

There are two additional sets of data used in the inference stage: a simulated gamma-ray dataset from Monte Carlo sampling and an off-region background dataset from real observations of the sky avoiding known sources (see Figure \ref{fig:off-source}). Monte Carlo simulations of background images from protons and electrons are computationally expensive, but background observations are a ``free" alternative. These simulated gamma-ray and observed background images become the reference for gamma and hadron signals, or an artificially labeled dataset for the inference stage. Throughout this paper, we use the real dataset, simulated dataset and background dataset to refer to the observed Crab Nebula dataset, simulated gamma-ray dataset, and the off-region background dataset.

Due to the wide distribution of image pixel values between noise- and EAS-dominated images, we used Fisher normalization for the training. The normalization is performed in an event-wise fashion, which allows us to normalize the data on the fly during training.
The process is to first normalize the dataset to the range $[-1,1]$ and then pass it to the \texttt{arctanh} operation. 
The Fisher normalization unskewed the original dataset, transforming a distribution of pixel values peaked at zero toward a more Gaussian-like distribution.



\section{WGAN-based Framework and Training Strategy} \label{sec:Experiment}
\begin{figure}[h]
    \centering
    \includegraphics[width=0.6\columnwidth]{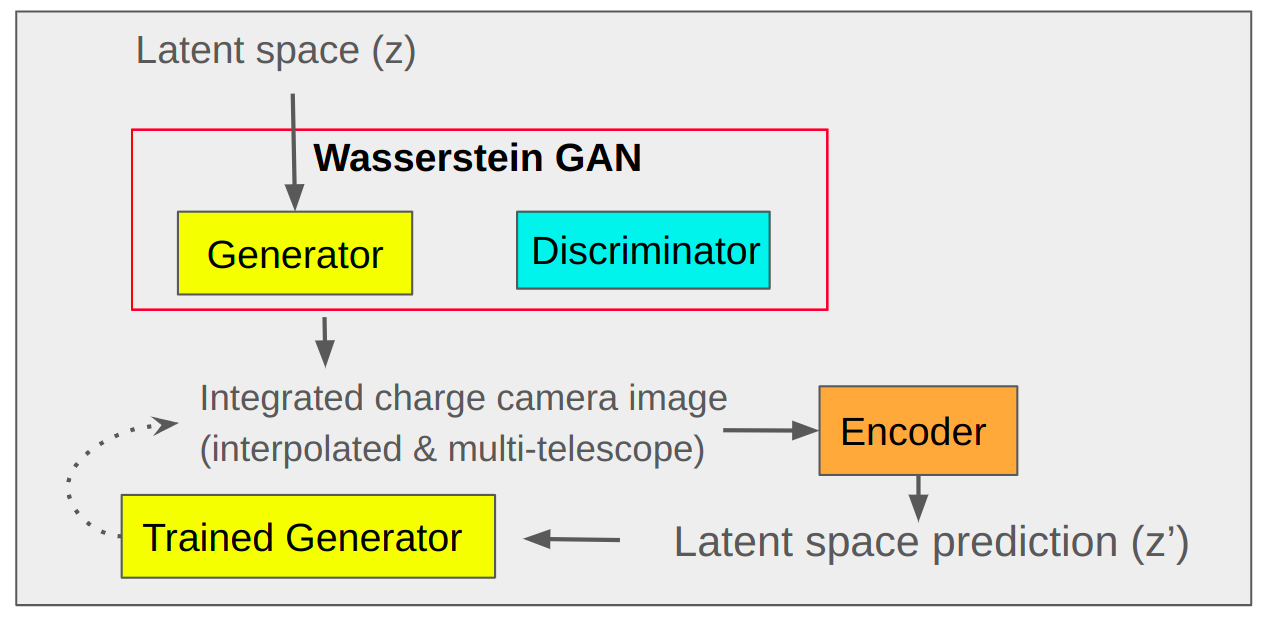}
    \caption{A WGAN-based model. The generator and discriminator components output generated images, while a separately trained encoder outputs latent space vectors. The combination of the GAN and encoder model enables direct comparison between a real image and its model-generated counterpart.}
    \label{fig:training_pipeline}
\end{figure}

The deep learning framework is inspired by the GAN-based anomaly score detection work \cite{schlegl2019f}. The framework consists of three neural network models: generator, discriminator and encoder. First, the generator and discriminator are trained simultaneously so that the generator can learn the feature space of the data and generate realistic data. Then, the encoder is trained to learn the inverse mapping from the images to the feature space. Therefore, our complete pipeline allows us to pass in observational data through the encoder and retrieve the corresponding latent vector which contains the feature representation. Figure \ref{fig:training_pipeline} shows the details of the WGAN-based model. During the GAN training, the generator maps the random latent vector to the image space to fool the discriminator.

\begin{figure}[h]
    \centering
    \includegraphics[width=0.9\columnwidth]{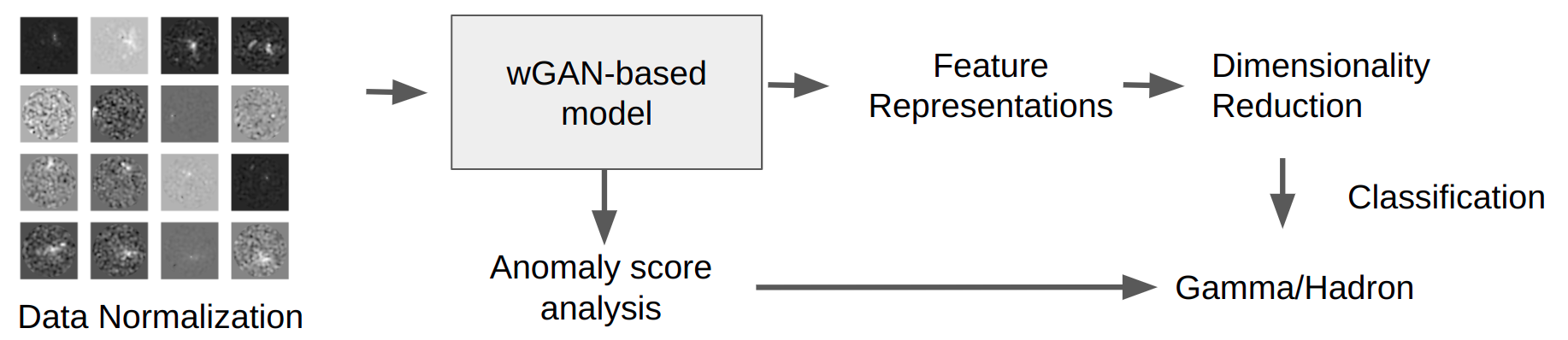}
    \caption{Pipeline of complete GAN-based framework. After model training, the feature representation is extracted and reduced for the inference and classification stage. Anomaly scores are also extracted from the trained model to support gamma/hadron classification.}
    \label{fig:Gan-based_framework}  
\end{figure}

The pipeline of our WGAN-based framework is shown in Figure \ref{fig:Gan-based_framework}. The WGAN model is trained on the stereoscopic integrated charge images and gives the feature representations of the image-based dataset in the form of latent vectors. The latent vectors are further embedded in lower dimensions via the dimensionality reduction methods. The classification of the dataset is conducted in the lower dimension space.

When the GAN training converges, we freeze the generator and discriminator to further train an encoder, which maps the images to the latent space and thus produces predicted latent vectors. The trained encoder is the bridge between real and reconstructed images, and provides the feature representations. The difference between the original images and the corresponding reconstructed images provides the anomaly scores, which are a proxy for the gamma-ray probability of an image. Conceptually, high anomaly scores correspond to those events that are poorly represented within the dataset, and we expect these to be preferentially gamma-like. We also further conduct dimensionality reduction through a combination of Principal Component Analysis(PCA) and Uniform Manifold Approximation and Projection (UMAP) algorithm on the latent vectors in preparation for the classification task \cite{mcinnes2018umap-software}.

For our experiments, the WGAN was trained for 750 epochs using an Adam optimizer ($\beta_{1} = 0.5$ and $\beta_{2} = 0.999$) with a learning rate of $5\times10^{-5}$ and a scheduled piecewise-constant decay\footnote{See \hyperlink{https://www.tensorflow.org/api_docs/python/tf/keras/optimizers/schedules/PiecewiseConstantDecay}{PiecewiseConstantDecay} for the Keras official document.} which halves the learning rate every 250 epochs. The gradient penalty coefficient is $\lambda = 10$. The encoder was trained for $500$ epochs also using an Adam optimizer ($\beta_{1} = 0.5$ and $\beta_{2} = 0.9$) with a constant learning rate $10^{-4}$.

\section{Results and Discussions}
\label{sec: Result and discussion}

\begin{figure}[t]
    \centering
    \subfigure
    {\includegraphics[width=0.4\textwidth]{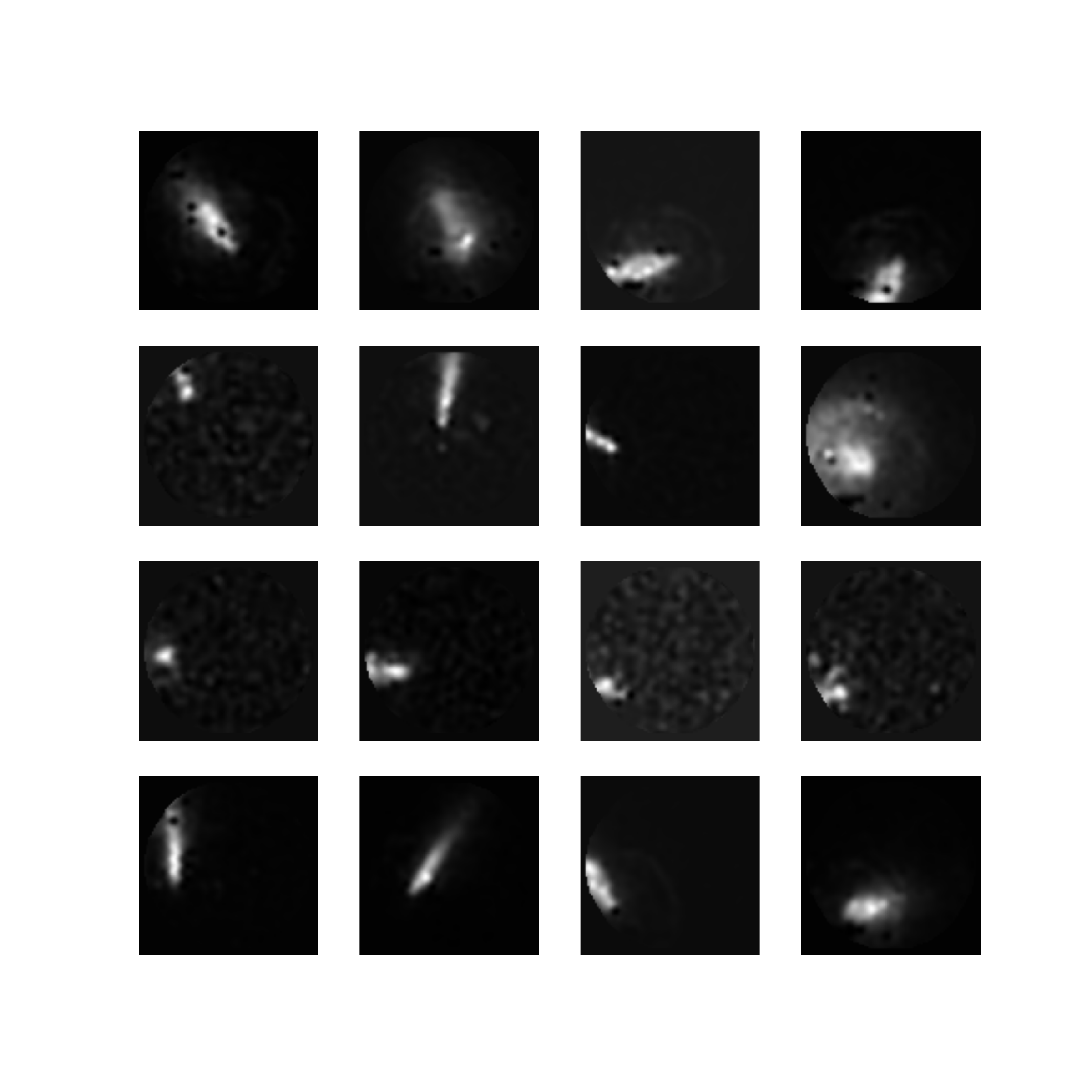}} 
    \subfigure{\includegraphics[width=0.4\textwidth]{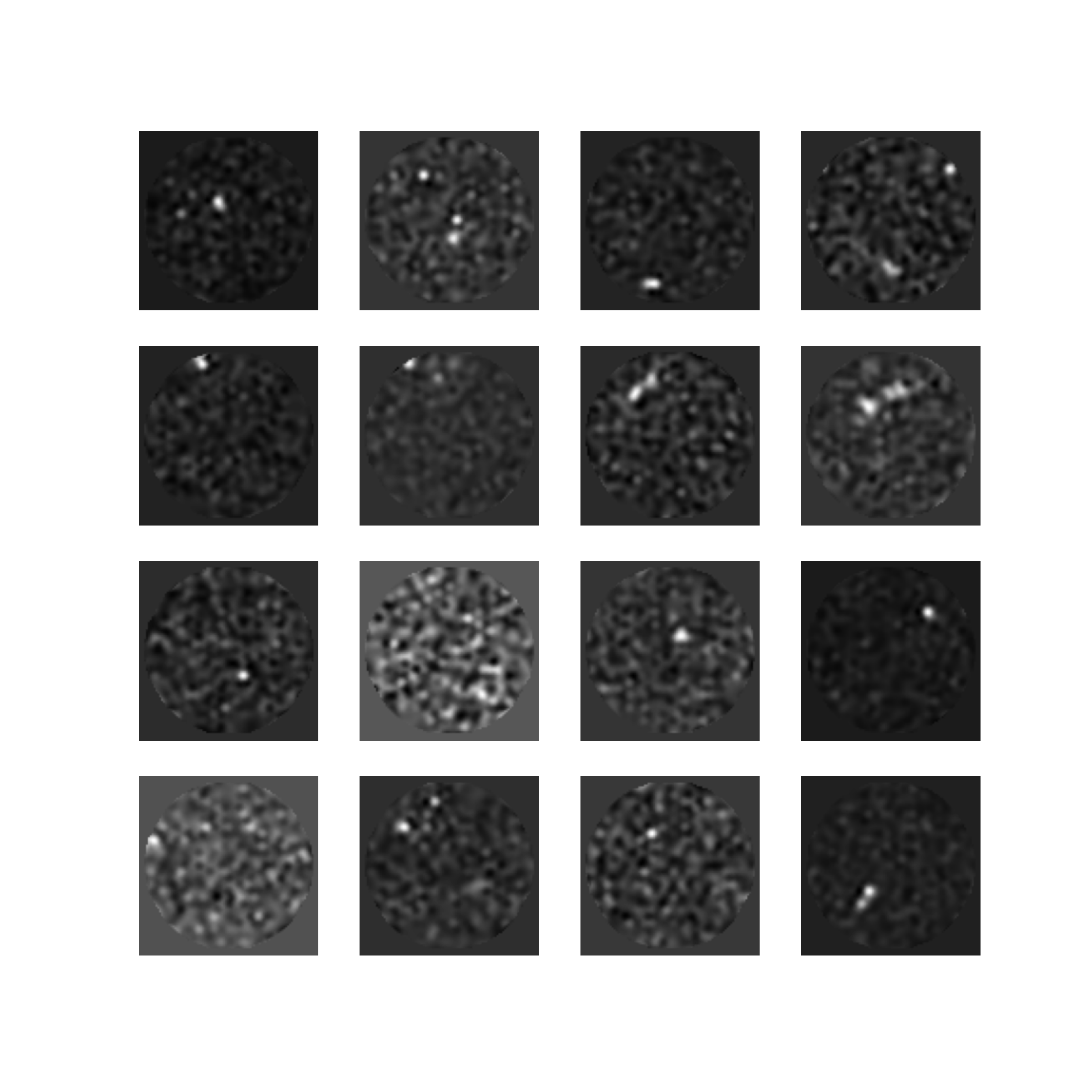}} 
    \vspace{-20pt}
    \caption{\textbf{Events with high and low anomaly scores. } The left plot shows 4 events (each row) with anomaly scores of 0.847, 0.936, 0.815 and 0.821 from top to bottom. The right plot shows 4 events (each row) with anomaly scores of 0.179, 0.195, 0.165, 0.188 from top to bottom. }
    \label{fig:anomaly_scores}
\end{figure}

\begin{figure}[h]
    \centering
    \subfigure
    {\includegraphics[width=0.24\textwidth]{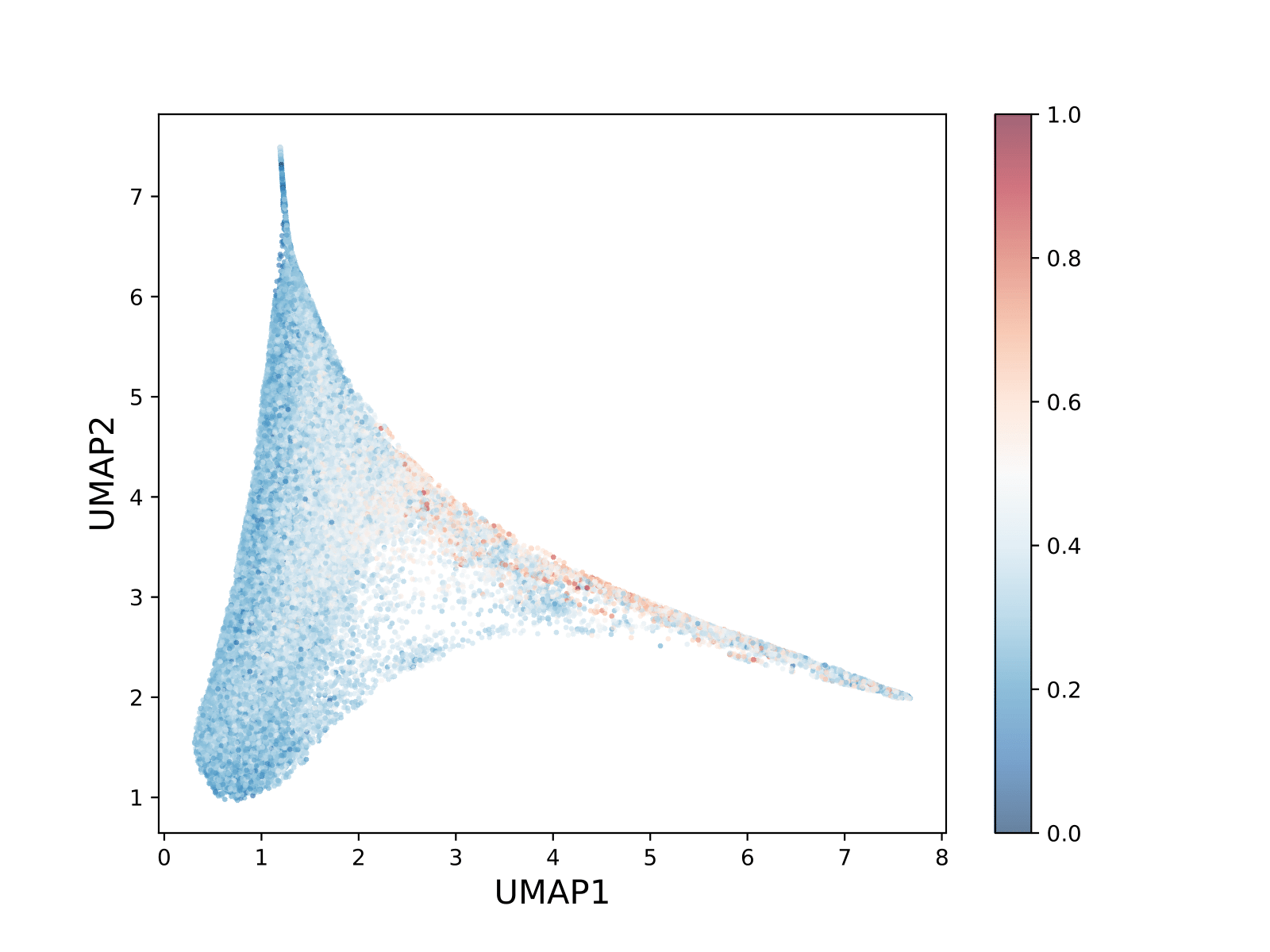}} 
    \subfigure{\includegraphics[width=0.24\textwidth]{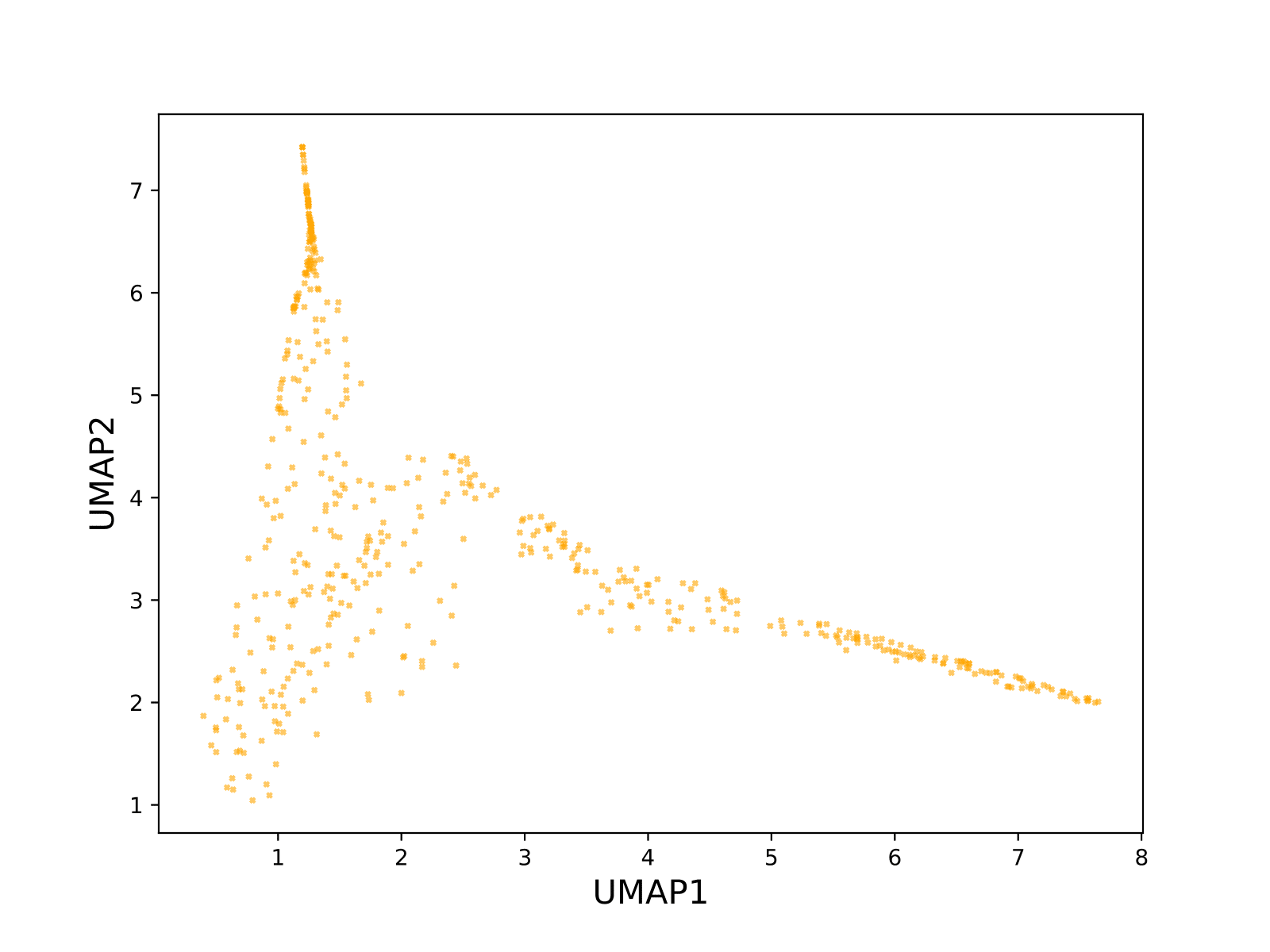}} 
    \subfigure{\includegraphics[width=0.24\textwidth]{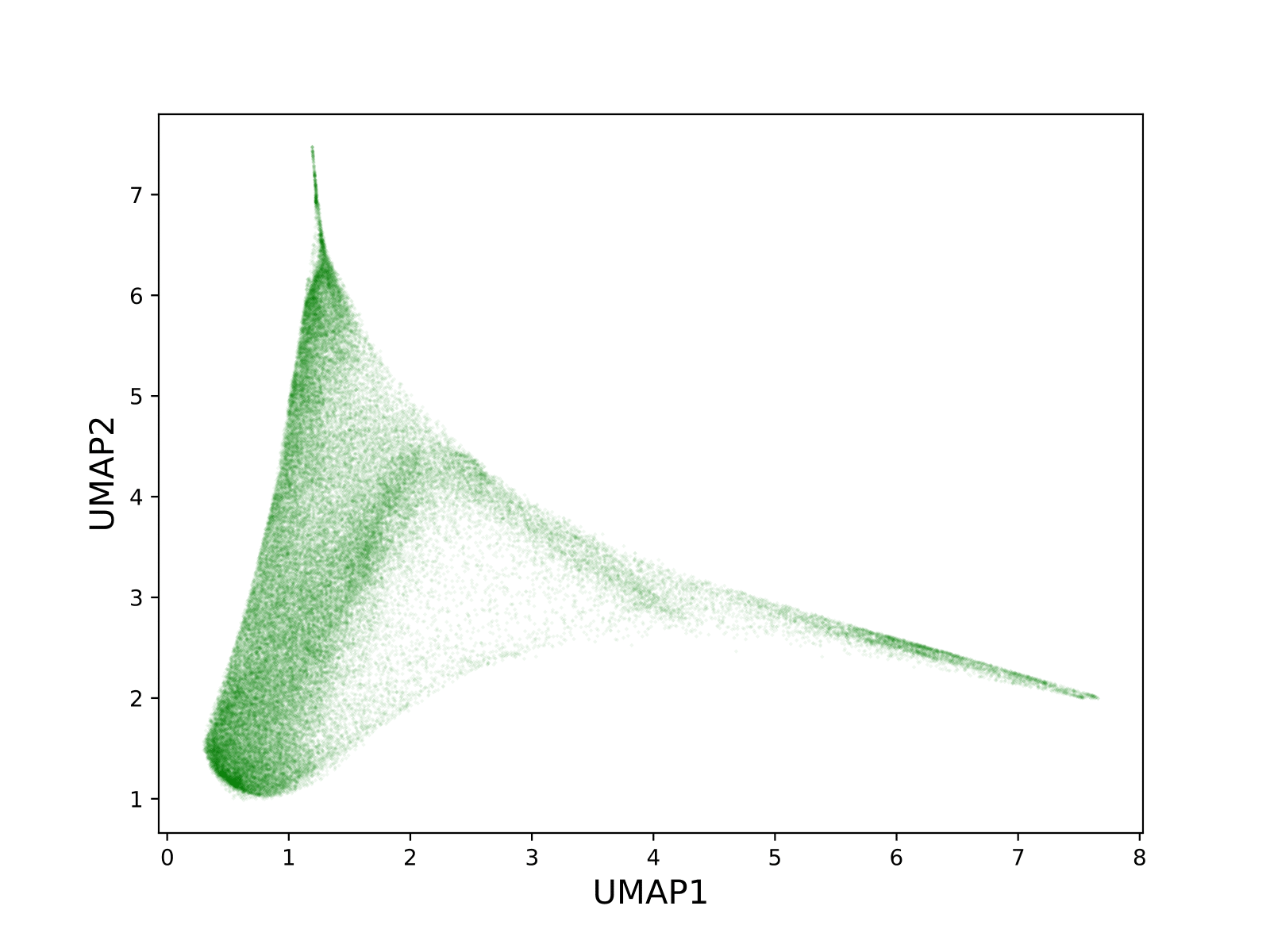}}

    \subfigure
    {\includegraphics[width=0.24\textwidth]{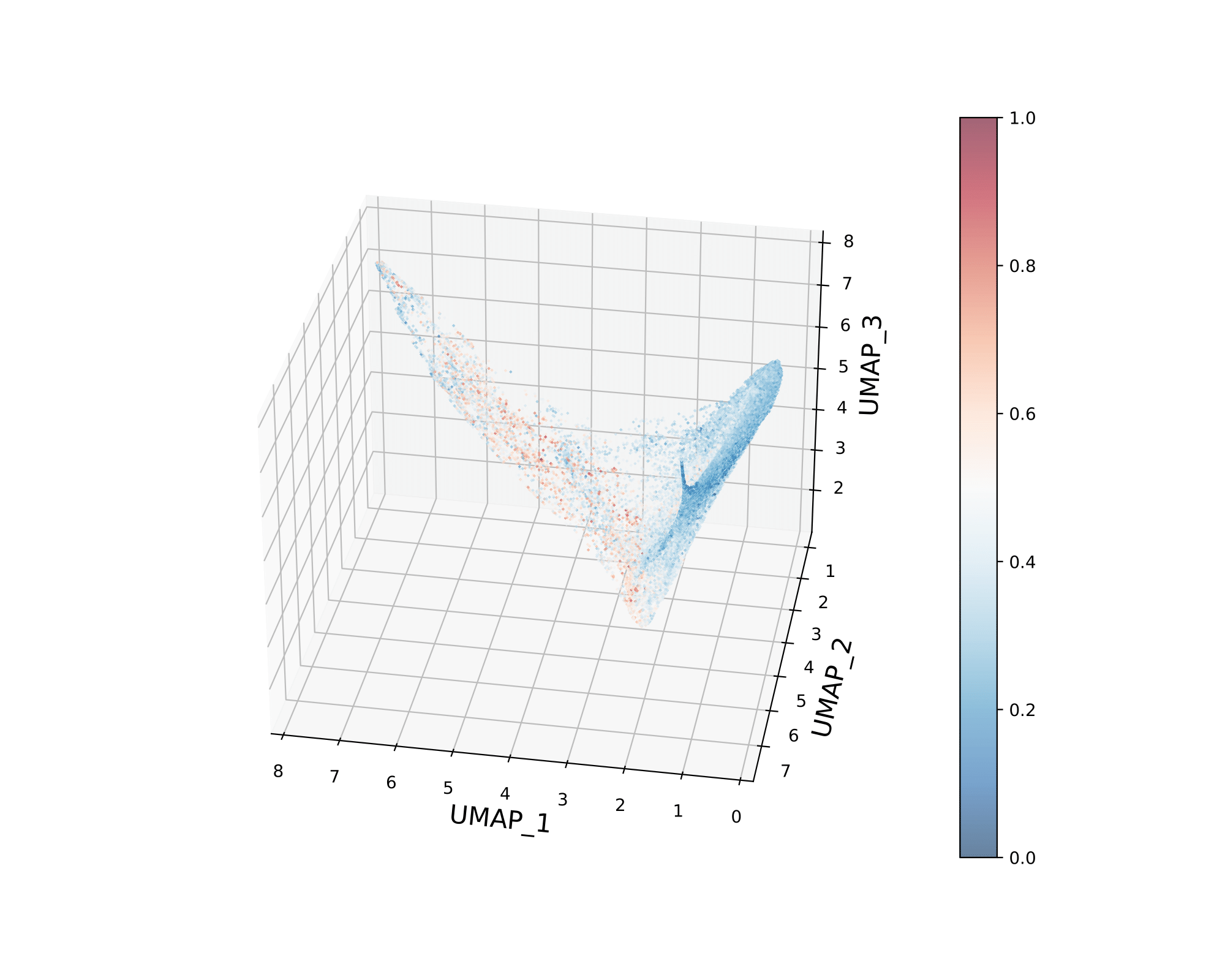}} 
    \subfigure{\includegraphics[width=0.24\textwidth]{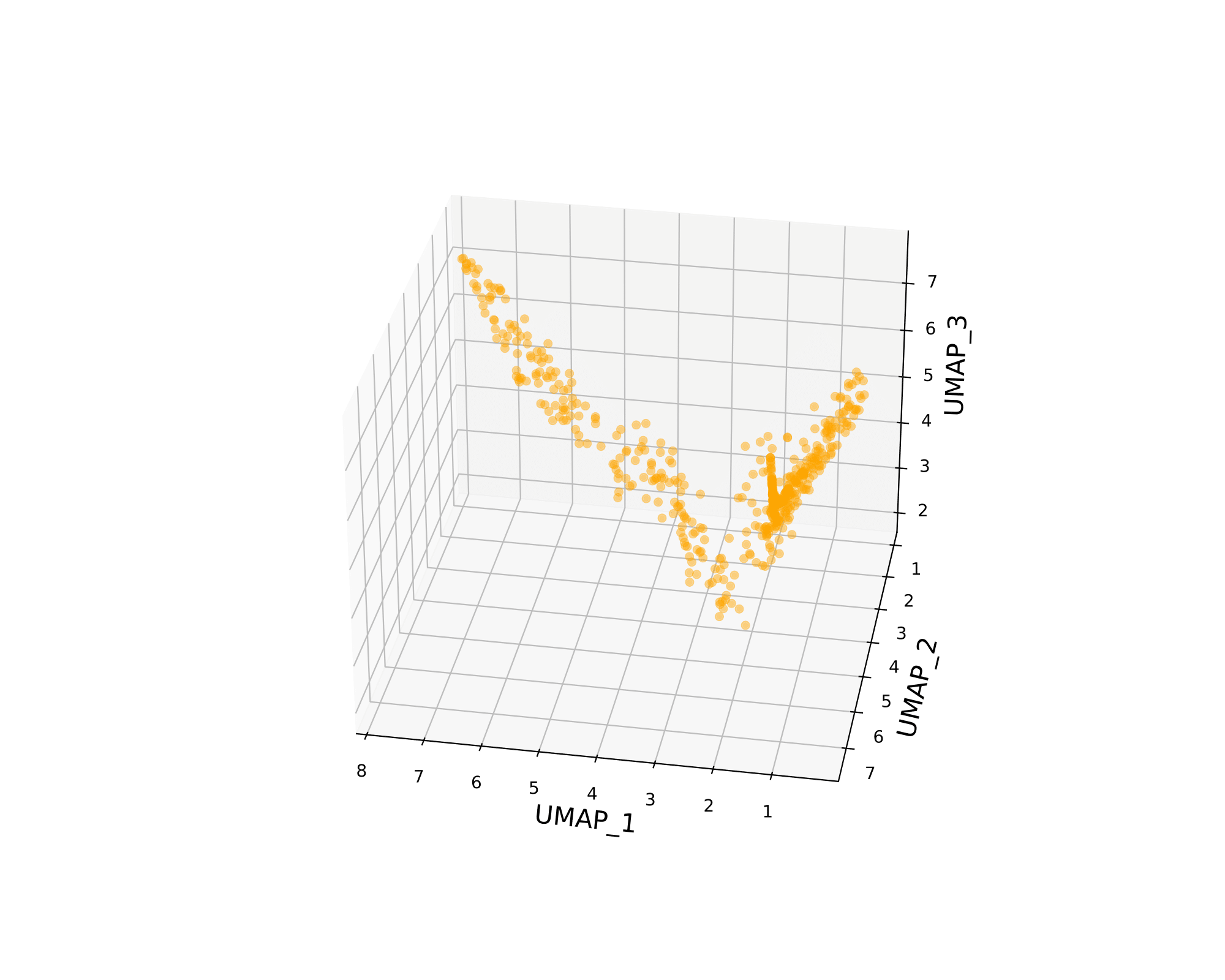}} 
    \subfigure{\includegraphics[width=0.24\textwidth]{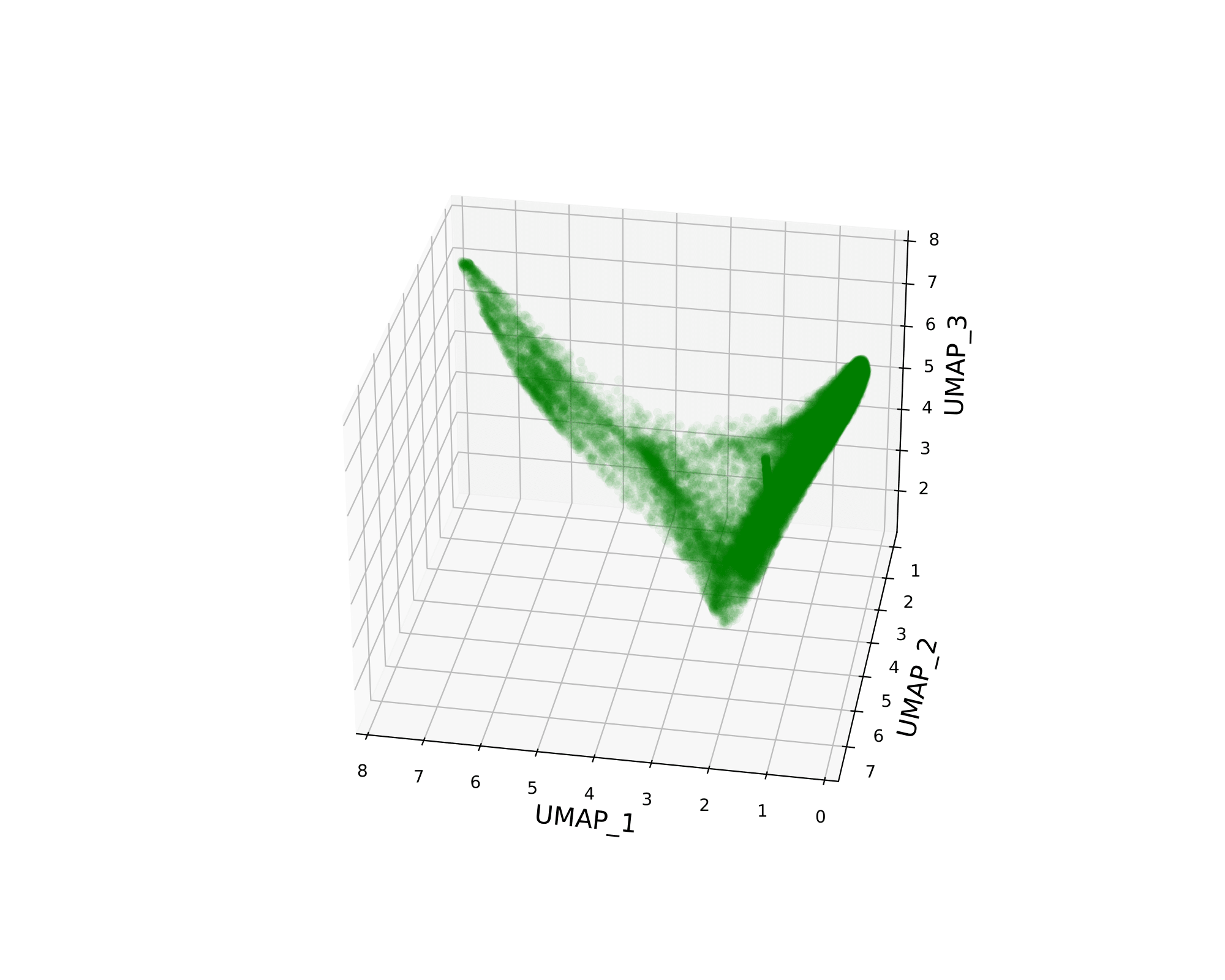}}
    \caption{\textbf{2D and 3D UMAPs}. First column is real data color-coded by anomaly scores. Encoded onto the same latent space, the second column is simulated data in yellow and the third column is the background data in green.}
    \label{fig:UMAPs}
\end{figure}

Figure \ref{fig:anomaly_scores} shows the generated images with high anomaly scores ($\geq 99.95$ percentile) and low anomaly scores ($\leq 2$ percentile). The high-anomaly-score images mostly have bright, elliptical signals, which have characteristics resembling EAS images. Meanwhile, the low-anomaly-score images have small dotted bright points in the camera plane, with a dominance of camera noise and may be corresponding to less-bright EAS signals. While the anomaly score trends offer some reasonable guidance WGAN feature learning, further verification is needed. As such, we assess the learned feature representations using UMAP of the real images, simulated images and background images, and investigate the distribution of the data as a function of the latent (feature) space (Figure \ref{fig:UMAPs}).

In the first column of Figure \ref{fig:UMAPs}, the low anomaly scores are distributed throughout the latent space, while a low density of high anomaly scores are scattered in the central region. The second and third columns of Figure \ref{fig:UMAPs} shows the derived latent space of the simulated and background data, respectively, which have very similar shapes and densities compared to the latent space of the original training data. In an ideal training result, we expect that the latent spaces of the simulated and background data overlap with the real data in different regions, showing that the real data in the latent space is separable into distinct gamma ray and background regions. Unfortunately, the latent spaces in Figure \ref{fig:UMAPs} are very similar and difficult to separate.

\begin{figure}[t]
    \centering
    \subfigure{\includegraphics[width=0.5\columnwidth]{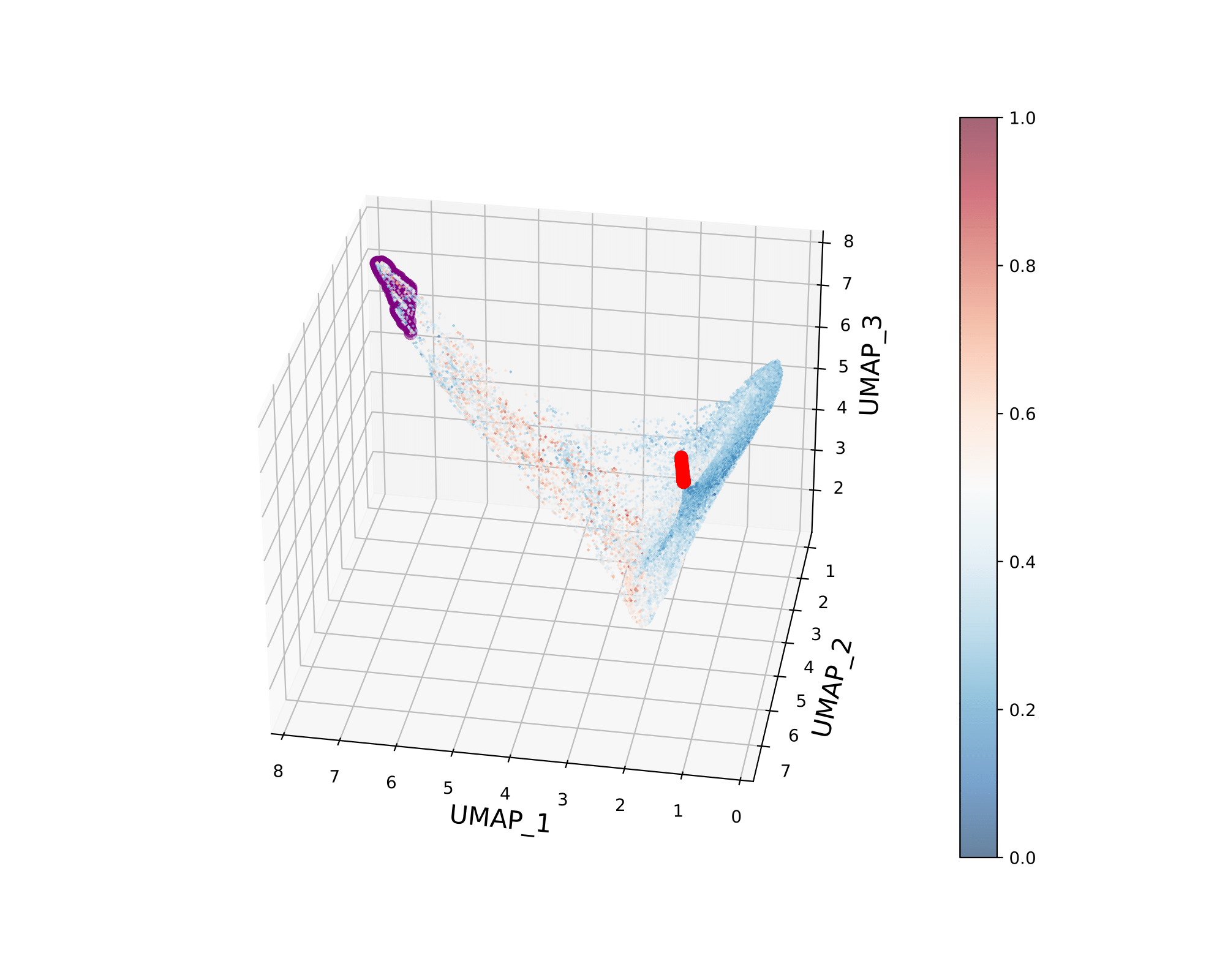}}
    \subfigure{
    \includegraphics[width=0.4\textwidth]{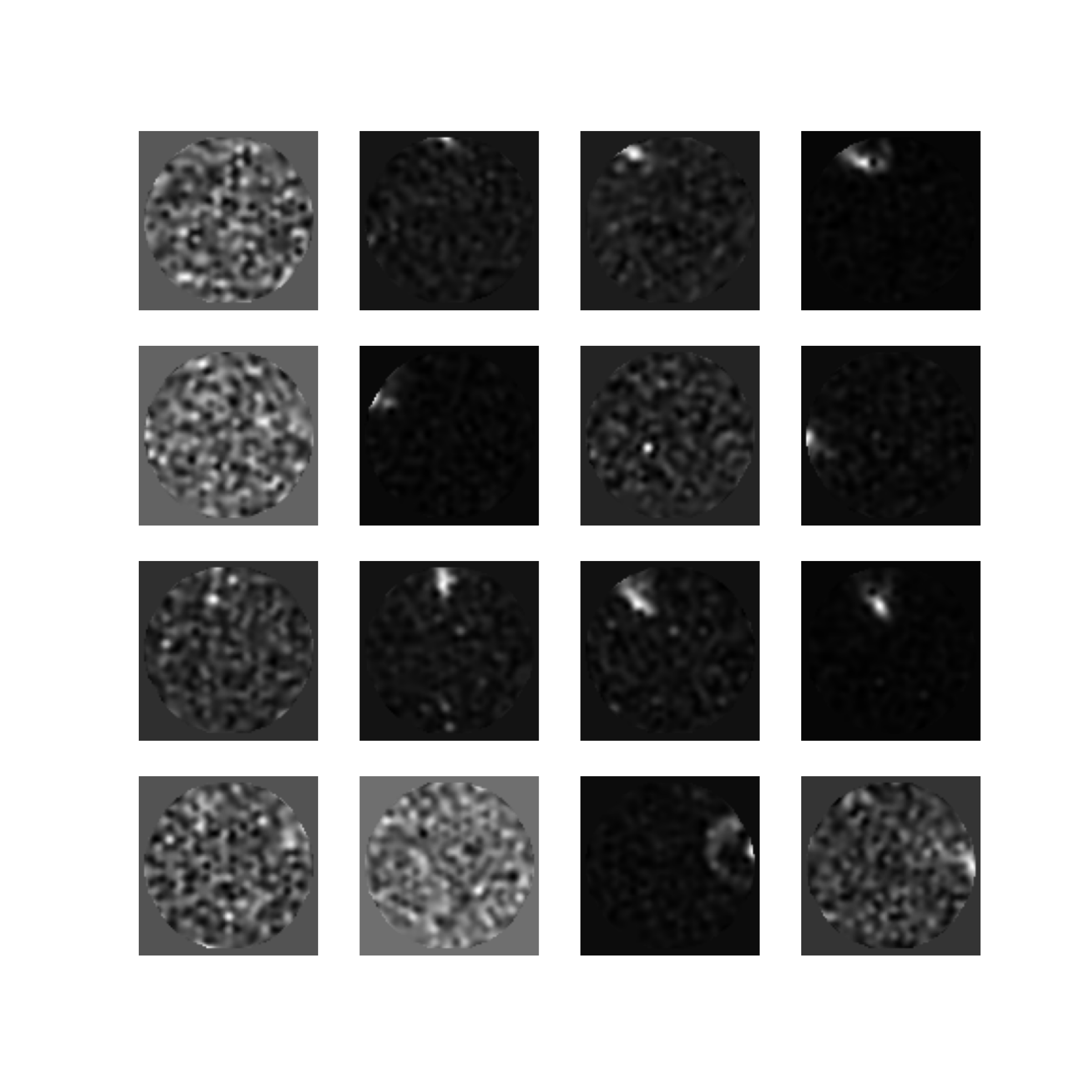}
    \includegraphics[width=0.4\textwidth]{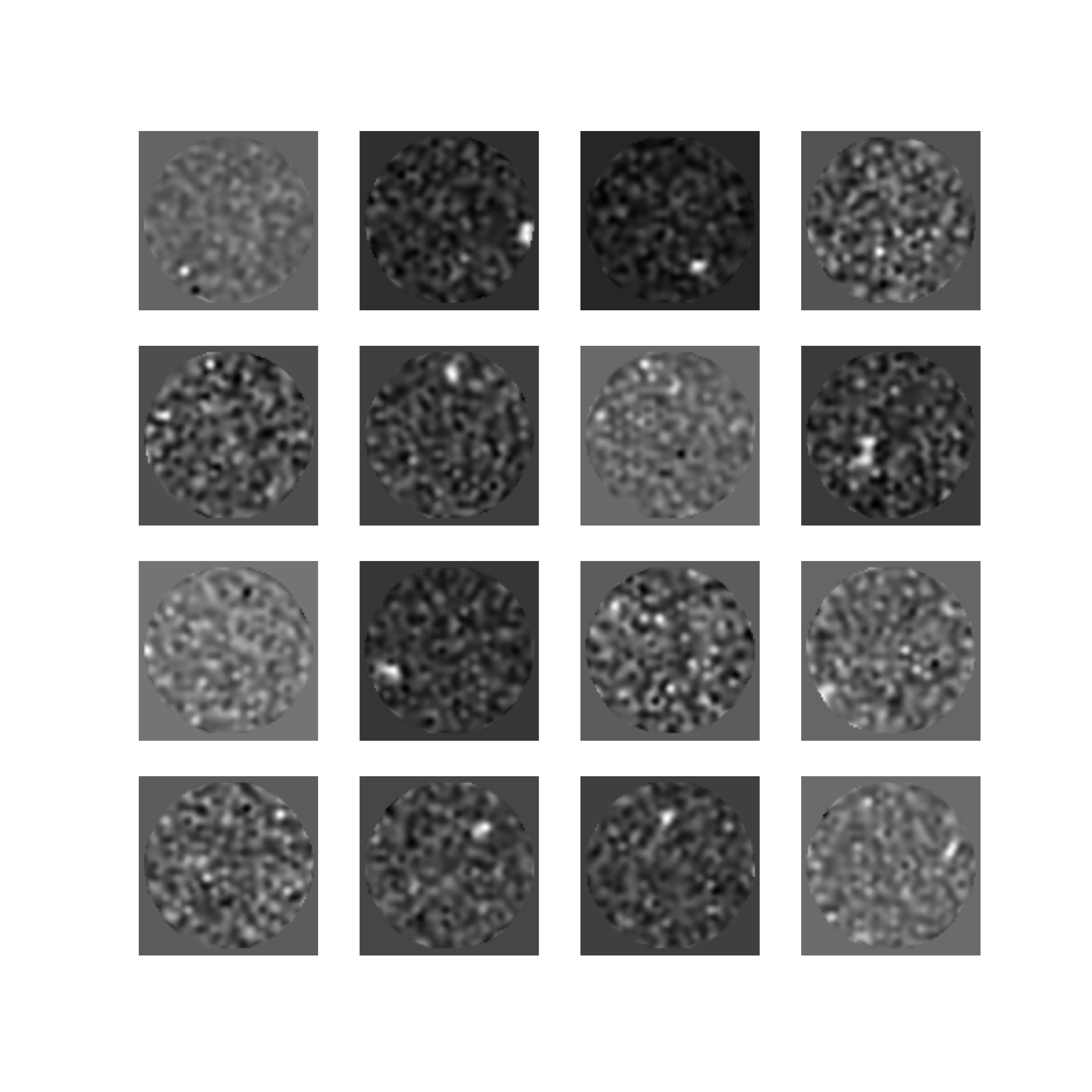}}
    \caption{\textbf{Top)} UMAP with interesting clusters highlighted in purple and red. \textbf{Bottom)} Images in the clusters.  The left images are from purple clusters above and have anomaly scores of 0.376, 0.370, 0.598 and 0.516 from top to bottom. The right images are from the red clusters above and have anomaly scores of 0.193, 0.157, 0.178, 0.278 from top to bottom. These images do not have the same color scales, so background dominated images appear brighter than signal dominated images.}
    \label{fig:anomaly_scores_images}
\end{figure}

To fully investigate the feature representations, we also observe the features of the off-centered data points shown in the UMAP. In Figure \ref{fig:anomaly_scores_images}, the purple and red points are two clusters of interest. We can see that the different clusters indeed have different features shown in the images. This is a good direction for exploring the separation of the hyperplane, but further refinement is needed to get a more convincing result.


In this work, we trained a WGAN-based anomaly detector and analyzed the latent feature space representations. Although our model achieved convergence, we find that the latent representations and anomaly scores do not align with our expectations of their correlations with the gamma vs. hadron differentiation. This suggests that our model has failed to learn generalized features that hold distinguishing power between gamma and hadronic signals. The model in turn seems to have learned a correlation with the relative signal strength in the images with respect to the noise (SNR), indicating that this may be because of the employed normalization scheme. As such, our future efforts will focus on implementing a more representative normalization scheme (e.g., generating normalized SNR images) for our model training and enforcing learning of normalized morphological features.

\section{Conclusion}
\label{sec:conclusion}
In this study we employ a data-driven unsupervised deep learning based approach using observational data from VERITAS towards gamma/hadron separability. We show that using a Wasserstein GAN architecture enables relatively successful model capable of generating EAS images. Although we could not complete the gamma/hadron separation problem within the dimensional reduction of the latent space, we show that further investigation of features in the latent space yields useful information for separating the hyperplane. Assessing images with interesting anomaly scores is a promising strategy for further experimentation. Modifications to the model are crucial to the issues revealed here, such as hyperparameter tuning and even increasing the dataset for training. The image normalization strategy is critical to learning the relevant features toward gamma/hadron separation, but is challenging in the context of IACT stereoscopic data. New strategies for a more robust normalization are being studied and are expected to yield solutions to the gamma/hadron separation problem.

\section{Acknowledgments}
This work was partially supported by NSF award PHY 2110737. This research is supported by grants from the U.S. Department of Energy Office of Science, the U.S. National Science Foundation and the Smithsonian Institution, by NSERC in Canada, and by the Helmholtz Association in Germany. This research used resources provided by the Open Science Grid, which is supported by the National Science Foundation and the U.S. Department of Energy's Office of Science, and resources of the National Energy Research Scientific Computing Center (NERSC), a U.S. Department of Energy Office of Science User Facility operated under Contract No. DE-AC02-05CH11231. We acknowledge the excellent work of the technical support staff at the Fred Lawrence Whipple Observatory and at the collaborating institutions in the construction and operation of the instrument.

\bibliographystyle{ICRC}
\bibliography{./ref}


%
%
%

\clearpage

\section*{Full Author List: VERITAS Collaboration}

\scriptsize
\noindent
A.~Acharyya$^{1}$,
C.~B.~Adams$^{2}$,
A.~Archer$^{3}$,
P.~Bangale$^{4}$,
J.~T.~Bartkoske$^{5}$,
P.~Batista$^{6}$,
W.~Benbow$^{7}$,
J.~L.~Christiansen$^{8}$,
A.~J.~Chromey$^{7}$,
A.~Duerr$^{5}$,
M.~Errando$^{9}$,
Q.~Feng$^{7}$,
G.~M.~Foote$^{4}$,
L.~Fortson$^{10}$,
A.~Furniss$^{11, 12}$,
W.~Hanlon$^{7}$,
O.~Hervet$^{12}$,
C.~E.~Hinrichs$^{7,13}$,
J.~Hoang$^{12}$,
J.~Holder$^{4}$,
Z.~Hughes$^{9}$,
T.~B.~Humensky$^{14,15}$,
W.~Jin$^{1}$,
M.~N.~Johnson$^{12}$,
M.~Kertzman$^{3}$,
M.~Kherlakian$^{6}$,
D.~Kieda$^{5}$,
T.~K.~Kleiner$^{6}$,
N.~Korzoun$^{4}$,
S.~Kumar$^{14}$,
M.~J.~Lang$^{16}$,
M.~Lundy$^{17}$,
G.~Maier$^{6}$,
C.~E~McGrath$^{18}$,
M.~J.~Millard$^{19}$,
C.~L.~Mooney$^{4}$,
P.~Moriarty$^{16}$,
R.~Mukherjee$^{20}$,
S.~O'Brien$^{17,21}$,
R.~A.~Ong$^{22}$,
N.~Park$^{23}$,
C.~Poggemann$^{8}$,
M.~Pohl$^{24,6}$,
E.~Pueschel$^{6}$,
J.~Quinn$^{18}$,
P.~L.~Rabinowitz$^{9}$,
K.~Ragan$^{17}$,
P.~T.~Reynolds$^{25}$,
D.~Ribeiro$^{10}$,
E.~Roache$^{7}$,
J.~L.~Ryan$^{22}$,
I.~Sadeh$^{6}$,
L.~Saha$^{7}$,
M.~Santander$^{1}$,
G.~H.~Sembroski$^{26}$,
R.~Shang$^{20}$,
M.~Splettstoesser$^{12}$,
A.~K.~Talluri$^{10}$,
J.~V.~Tucci$^{27}$,
V.~V.~Vassiliev$^{22}$,
A.~Weinstein$^{28}$,
D.~A.~Williams$^{12}$,
S.~L.~Wong$^{17}$,
and
J.~Woo$^{29}$\\
\\
\noindent
$^{1}${Department of Physics and Astronomy, University of Alabama, Tuscaloosa, AL 35487, USA}

\noindent
$^{2}${Physics Department, Columbia University, New York, NY 10027, USA}

\noindent
$^{3}${Department of Physics and Astronomy, DePauw University, Greencastle, IN 46135-0037, USA}

\noindent
$^{4}${Department of Physics and Astronomy and the Bartol Research Institute, University of Delaware, Newark, DE 19716, USA}

\noindent
$^{5}${Department of Physics and Astronomy, University of Utah, Salt Lake City, UT 84112, USA}

\noindent
$^{6}${DESY, Platanenallee 6, 15738 Zeuthen, Germany}

\noindent
$^{7}${Center for Astrophysics $|$ Harvard \& Smithsonian, Cambridge, MA 02138, USA}

\noindent
$^{8}${Physics Department, California Polytechnic State University, San Luis Obispo, CA 94307, USA}

\noindent
$^{9}${Department of Physics, Washington University, St. Louis, MO 63130, USA}

\noindent
$^{10}${School of Physics and Astronomy, University of Minnesota, Minneapolis, MN 55455, USA}

\noindent
$^{11}${Department of Physics, California State University - East Bay, Hayward, CA 94542, USA}

\noindent
$^{12}${Santa Cruz Institute for Particle Physics and Department of Physics, University of California, Santa Cruz, CA 95064, USA}

\noindent
$^{13}${Department of Physics and Astronomy, Dartmouth College, 6127 Wilder Laboratory, Hanover, NH 03755 USA}

\noindent
$^{14}${Department of Physics, University of Maryland, College Park, MD, USA }

\noindent
$^{15}${NASA GSFC, Greenbelt, MD 20771, USA}

\noindent
$^{16}${School of Natural Sciences, University of Galway, University Road, Galway, H91 TK33, Ireland}

\noindent
$^{17}${Physics Department, McGill University, Montreal, QC H3A 2T8, Canada}

\noindent
$^{18}${School of Physics, University College Dublin, Belfield, Dublin 4, Ireland}

\noindent
$^{19}${Department of Physics and Astronomy, University of Iowa, Van Allen Hall, Iowa City, IA 52242, USA}

\noindent
$^{20}${Department of Physics and Astronomy, Barnard College, Columbia University, NY 10027, USA}

\noindent
$^{21}${ Arthur B. McDonald Canadian Astroparticle Physics Research Institute, 64 Bader Lane, Queen's University, Kingston, ON Canada, K7L 3N6}

\noindent
$^{22}${Department of Physics and Astronomy, University of California, Los Angeles, CA 90095, USA}

\noindent
$^{23}${Department of Physics, Engineering Physics and Astronomy, Queen's University, Kingston, ON K7L 3N6, Canada}

\noindent
$^{24}${Institute of Physics and Astronomy, University of Potsdam, 14476 Potsdam-Golm, Germany}

\noindent
$^{25}${Department of Physical Sciences, Munster Technological University, Bishopstown, Cork, T12 P928, Ireland}

\noindent
$^{26}${Department of Physics and Astronomy, Purdue University, West Lafayette, IN 47907, USA}

\noindent
$^{27}${Department of Physics, Indiana University-Purdue University Indianapolis, Indianapolis, IN 46202, USA}

\noindent
$^{28}${Department of Physics and Astronomy, Iowa State University, Ames, IA 50011, USA}

\noindent
$^{29}${Columbia Astrophysics Laboratory, Columbia University, New York, NY 10027, USA}

\end{document}